\documentclass[10pt]{iopart}

\usepackage{graphicx}
\usepackage{subfigure}
\usepackage{verbatim}
\usepackage{dcolumn}
\usepackage{bm}
\usepackage{epsf}
\usepackage{color}
\usepackage[colorlinks=true,citecolor=blue,linkcolor=blue,urlcolor=blue]{hyperref}
\usepackage{hhline}
\usepackage{amssymb}
\usepackage{iopams}
\usepackage{cite}
\usepackage{tikz}
\usetikzlibrary{arrows,shapes,calc,matrix}

  \expandafter\let\csname equation*\endcsname\relax
  \expandafter\let\csname endequation*\endcsname\relax
\usepackage{amsmath}

\makeatletter
\newcommand\footnoteref[1]{\protected@xdef\@thefnmark{\ref{#1}}\@footnotemark}
\makeatother

\newcommand{\bra}[1]{\left\langle #1\right|}
\newcommand{\ket}[1]{\left|#1\right\rangle}

\newcommand{\trace}[1]{\mathrm{tr}\left\{#1\right\}}

\newcommand{\la}{\left\langle}
\newcommand{\ra}{\right\rangle}
\newcommand{\pd}{\partial}

\newcommand{\mi}[1]{\min{\left\{#1\right\}}}
\newcommand{\ex}[1]{\exp{\left(#1\right)}}

\newcommand{\com}[2]{\left[#1,\,#2\right]}

\newcommand{\co}[1]{\cos{\left(#1\right)}}
\newcommand{\si}[1]{\sin{\left(#1\right)}}

\newcommand{\bla}{bla\\bla\\bla\\bla\\bla}

\newcommand{\mc}[1]{\mathcal{#1}}

\newcommand{\mrm}[1]{\mathrm{#1}}

\newcommand{\draftmode}{1}    
\newcommand{\notetoself}[1]{\ifnum \draftmode=1 {\color[rgb]{0,0,0.8} [#1]} \fi}  
\newcommand{\cuttext}[1]{\ifnum \draftmode=1 {\color[rgb]{0,0.5,0} [#1]} \fi}  
\newcommand{\warntext}[1]{\ifnum \draftmode=1 {\color[rgb]{0.9,0.6,0} #1} \else {#1} \color{black} \fi}

\begin{document}

\title{Geometric quantum speed limits: a case for Wigner phase space}

\author{Sebastian Deffner}
\address{Department of Physics, University of Maryland Baltimore County, Baltimore, MD 21250, USA}

\date{}
\ead{deffner@umbc.edu}

\begin{abstract}
The quantum speed limit is a fundamental upper bound on the speed of quantum evolution. However, the actual mathematical expression of this fundamental limit depends on the choice of a measure of distinguishability of quantum states. We show that quantum speed limits are universally characterized by the Schatten-$p$-norm of the generator of quantum dynamics. Since computing Schatten-$p$-norms  can be mathematically involved, we then develop an alternative approach in Wigner phase space. We find that the quantum speed limit  in Wigner space is fully equivalent to expressions in density operator space, but that the new bound is significantly easier to compute. Our results are illustrated for the parametric harmonic oscillator and for quantum Brownian motion.
\end{abstract}

\section{Introduction}

It has recently been argued that already the first generation of real-life quantum computers will be able to perform certain tasks exponentially faster than classical computers \cite{Boixo2016}. This so-called ``quantum supremacy'' \cite{Preskill2012} rests in the fact that loosely speaking quantum state space is exponentially larger than the classical computational space, and hence significantly less operations are necessary to perform the same computation. However, the working principles of quantum computers and classical computers are fundamentally different, which makes it not immediately clear how to quantify the ``quantum speed-up'' \cite{Ronnow2014}. To make matters even more involved, in the theory of quantum computation ``time'' is not actually a  physical time, but rather a synonym for the ``number of computational operations'' \cite{Ronnow2014}. The more practical question is, however, how fast a quantum computer could actually operate.

To address this issue a somewhat opposite approach has been developed in quantum dynamics, where the notion of a \emph{quantum speed limit} has found wide-spread prominence. Whereas in the theory of quantum computation one is after characterizing quantum speed-ups -- the quest for faster and faster computations with less and less single operations -- the quantum speed limit sets the ultimate, maximal speed with which any quantum system can evolve. This means, in particular, that every single quantum operation takes a finite, minimal time to be accomplished -- and thus even quantum computers will not be able to achieve any arbitrary speed-ups. This quantum speed limit originates in the Heisenberg indeterminacy principle\cite{Heisenberg1927,Heisenberg2008}, $\Delta E\,\Delta t\gtrsim \hbar/2$. However, more than juts being another expression of  quantum indeterminacy  the quantum speed limit is a  fundamental property of quantum dynamics as highlighted by famous debates between Einstein and Bohr \cite{Hilgevoord1998}.

The first rigorous treatment was propsoed by Mandelstam and Tamm \cite{mandelstam45}, who showed that the minimal time a quantum system needs to evolve between orthogonal states is bounded from below by the variance of the energy, $\tau_\mrm{QSL}=\pi\hbar/2\Delta E$, where $\Delta E=(\la H^2\ra-\la H^2\ra)^{1/2}$. Since, however, the variance of an operator is not necessarily a good quantifier for dynamics \cite{Uffink1993}, Margolus and Levitin \cite{margolus98} revisited the problem and derived a second bound on the quantum evolution time in terms of the average energy $E=\la H\ra-E_g$ over the ground state with energy $E_g$, $\tau_\mrm{QSL}=\pi\hbar/2E$. It was eventually realized that these two bounds are not independent, and that only the unified bound is tight \cite{Levitin2009}.

Nowadays, it has been established in virtually all areas of quantum physics \cite{Bekenstein1981,lloyd00,Deffner2010,Giovannetti2011,Santos2015,Campbell2017} that the quantum speed limit \cite{Bhattacharyya1983,Pfeifer1993,Giovannetti2004,Frey2016}, sets a fundamental upper bound on the speed of any quantum dynamics.  In particular in  recent years, the quantum speed limit has been extensively studied and generalized for isolated \cite{Barnes2013a,Poggi2013,Hegerfeldt2013,Andersson2014,Deffner2013,richerme14} and open \cite{Deffner2013PRL,delcampo13,taddei13,deffner14,Zhang2014,Mukherjee2013,Xu2014,Xu2014a,Cimmarusti2015,Marvian2015,Hu2015,Mondal2016,Marvian2016} quantum systems. The renewed and concentrated interest was inspired by three letters \cite{Deffner2013PRL,delcampo13,taddei13}, which broadened the scope of  the quantum speed limit beyond unitary dynamics. In particular, Ref.~\cite{Deffner2013PRL} showed that the maximal speed of quantum evolution is given by the time averaged norm of the generator of the dynamics \cite{Deffner2013PRL}, which only in the case of unitary dynamics and for orthogonal states reduces to the average energy $E$ \cite{Deffner2013}. 

All treatments of the quantum speed limit have in common that the analyses start with a \emph{choice} of a measure of the distinguishability of quantum states. For instance, Ref.~\cite{delcampo13} studies the relative purity, Ref.~\cite{Deffner2013PRL} starts with the Bures angle between initially pure states and time-evolved states, and Ref.~\cite{Mondal2016PLA} even defines an entirely new metric. Finally, Pires \etal \cite{Pires2016} derived a whole family of quantum speed limits, which is based on a family of contractive Riemannian metrics. 

Thus, two natural questions arise: (i) Are all of these treatments of quantum speed independent, or can the quantum speed limit be universally characterized -- independently of the chosen measure of distinguishability? (ii) Most of the expressions for the quantum speed limit are mathematically rather involved. Hence, can one find a mathematically simple bound that is computable and nevertheless captures the universal behavior? 

In the following we will answer both questions. First, we will show that the quantum speed is universally characterized by any Schatten-$p$-norm of the generator of the quantum dynamics. In the second part of the analysis, we will derive a new quantum speed limit in terms of the Wasserstein norm of the rate of change of the Wigner function. We will argue that the quantum speed limit in Wigner phase space captures the same qualitative behavior as the speed limits derived in density operator space. However, we will also see that the new quantum speed limit is significantly easier to compute, for pure as well as mixed states, and for isolated as well as open dynamics. As an illustrative example we will discuss the semi-classical, high temperature limit, and we will confirm that the quantum speed limit is a pure quantum feature, i.e., that classical systems do not experience a fundamental bound on their rates of change.

\section{Quantum speed and the geometric approach}

We begin by briefly reviewing the main results of Ref.~\cite{Deffner2013PRL} and by establishing notions and notations. Consider a quantum master equation,
\begin{equation}
\label{eq01}
\dot\rho_t=L_t(\rho_t)\,,
\end{equation}
where the dot denotes a derivative with respect to time. For isolated systems the Liouvillian superoperator, $L_t$, reduces to the von-Neumann equation, $L_t(\rho)=\com{H_t}{\rho_t}/i\hbar$, but we explicitly allow for any open systems dynamics -- Markovian as well as non-Markovian. Note that Ref.~\cite{Deffner2013PRL} theoretically predicted that non-Markovian environments can speed up quantum dynamics. This was experimentally verified in cavity QED \cite{Cimmarusti2015}. 

In geometric quantum mechanics \cite{Bengtsson2007} it has proven useful to quantify the distinguishability of quantum states in terms of the Bures angle \cite{Bures1969},
\begin{equation}
\label{eq02}
\mc{L}(\rho_0,\rho_t)=\arccos\left(\sqrt{F(\rho_0,\rho_t)}\right)=\arccos\left(\trace{\sqrt{\sqrt{\rho_0}\,\rho_t\,\sqrt{\rho_0}}}\right)
\end{equation}
where we further introduced the quantum fidelity $F(\rho_0,\rho_t)$ \cite{Josza1994}. To obtain an upper bound on the speed of evolution one then considers the magnitude of the geometric speed, $|\dot{\mc{L}}|$, and it is easy to see that we have \cite{Deffner2013PRL}
\begin{equation}
\label{eq03}
2\co{\mc{L}}\si{\mc{L}}\,\dot{\mc{L}}\leq \left|\dot{F}(\rho_0,\rho_t)\right|\,.
\end{equation}
For initially pure states, $\rho_0=\ket{\psi_0}\bra{\psi_0}$, Eq.~\eqref{eq02} can be further simplified and it can be shown that \cite{Deffner2013PRL}
\begin{equation}
\label{eq04}
2\co{\mc{L}}\si{\mc{L}}\,\dot{\mc{L}}\leq \left|\bra{\psi_0}\dot{\rho}_t\ket{\psi_0}\right|\leq\mi{\|\dot{\rho}_t\|_p \,\,\mrm{for}\,\, p\in\{1,2,\infty\}}\,.
\end{equation}
Here $\|A\|_p$ denotes the Schatten-$p$-norm of an operator, $O$, which is defined as
\begin{equation}
\label{eq05}
\|O\|_p\equiv \left(\trace{\left|O\right|^p}\right)^{1/p}=\left(\sum_k o_k^p\right)^{1/p}
\end{equation}
and the $o_k$ are the singular values of $O$, i.e. the eigenvalues of the Hermitian operator $|O|\equiv\sqrt{O^\dagger O}$. The more familiar trace, Hilbert-Schmidt, and operator norms correspond respectively to $p=1,\, 2,$ and $\infty$. Equation~\eqref{eq04} can then be used to define the quantum speed limit $v_\mrm{QSL}$,
\begin{equation}
\label{eq05a}
\dot{\mc{L}}\leq \frac{v_\mrm{QSL}}{2\co{\mc{L}}\si{\mc{L}}}\equiv \frac{1}{2\co{\mc{L}}\si{\mc{L}}}\mi{\|\dot{\rho}_t\|_p \,\,\mrm{for}\,\, p\in\{1,2,\infty\}}\,.
\end{equation}
Note that in contrast to Ref.~\cite{Campbell2017} we did not include the denominator into the definition of $v_\mrm{QSL}$. The reason for this choice will become obvious shortly.

Although useful for theoretcial predictions of experimental outcomes \cite{Cimmarusti2015} Eq.~\eqref{eq04} also left several questions unaddressed. Probably the most immediate one is, how the above treatment would have to be generalized to initially mixed states. Generally this is a mathematically involved problem, since the quantum fidelity, $F(\rho_0,\rho_t)$, and its derivatives are non-trivial to handle. A comprehensive analysis of this issue was proposed by Pires \etal \cite{Pires2016}. They showed that if one considers the angle defined in terms of the Wigner-Yanase information, $A(\rho_0,\rho_t)$, 
\begin{equation}
\label{eq06}
\mc{L}^\mrm{WY}(\rho_0,\rho_t)=\arccos\left(A(\rho_0,\rho_t)\right)=\arccos\left(\trace{\sqrt{\rho_0}\sqrt{\rho_t}}\right)\,,
\end{equation}
instead of the Bures angle \eqref{eq02}, then an infinite family of bounds of the quantum speed can be derived. However, similarly to Eq.~\eqref{eq05a} the quantum speed is bounded by a norm of the generator of the dynamics, $\|\dot{\rho}_t\|$, and only the ``prefactor'' depends on the choice of the starting point, whether it be Eq.~\eqref{eq02} or Eq.~\eqref{eq06}. Thus, the analysis of Ref.~\cite{Pires2016} makes the second question even more obvious: Namely, formulating quantum speed limits seems to be somewhat arbitrary, since every single treatment starts with a \emph{choice} of a measure of distinguishability in order to define the geometric speed. Hence, it would be desirable to find a universal measure of quantum speed, and define the quantum speed limit exclusively in terms of this measure.

Both, Ref.~\cite{Deffner2013PRL} and Ref.~\cite{Pires2016} make the strong case that the choice has to be a contractive, Riemannian metric on quantum state space. This choice is justified if one would like to find tight bounds on the quantum speed \cite{Deffner2013PRL}. If one is only interested in the qualitative behavior, however, other measures might be more convenient to work with. For instance, instead of choosing the Bures angle, $\mc{L}(\rho_0,\rho_t)$, we also could have worked with the Bures distance \cite{Bengtsson2007},
\begin{equation}
\label{eq07}
\mc{L}^D(\rho_0,\rho_t)=\sqrt{2\left(1-\sqrt{F(\rho_0,\rho_t)}\right)}\,.
\end{equation}
One easily convinces oneself that a such defined geometric speed, $\dot{\mc{L}}^D$, leads to an inequality similar to Eq.~\eqref{eq03}. Therefore, we observe already here  that all of these treatments have in common that eventually the quantum speed is characterized by a Schatten-$p$-norm of the generator of the dynamics, $\|\dot{\rho_t}\|_p=\| L(\rho_t)\|_p$.

\section{Quantum speed in density operator space}

We have seen above that typically quantum speed is characterized by the dynamical behavior of the quantum fidelity, $F(\rho_0,\rho_t)$. Since $F(\rho_0,\rho_t)$ is mathematically rather involved several continuity bounds have been derived. For instance, we have \cite{Bengtsson2007}
\begin{equation}
\label{eq08}
1-\sqrt{F(\rho_0,\rho_t)}\leq \frac{1}{2}\ell_1(\rho_t,\rho_0)\leq \sqrt{1-F(\rho_0,\rho_t)}\,,
\end{equation}
where $\ell_1$ denotes the trace distance, i.e., the Schatten-1-distance
\begin{equation}
\label{eq09}
\ell_1(\rho_t,\rho_0)=\|\,\rho_t-\rho_0\,\|_1\equiv\trace{\left|\rho_t-\rho_0\right|}\,.
\end{equation}
The obvious question is whether a quantum speed limit can be derived starting with the trace distance $\ell_1$. To this end, we consider the geometric speed
\begin{equation}
\label{eq10}
\dot{\ell}_1(\rho_t,\rho_0)=\trace{\left|\rho_t-\rho_0\right|^{-1}\,\left(\rho_t-\rho_0\right)\,\dot{\rho}_t}\,,
\end{equation}
which can be bounded from above by with the triangle inequality for operators, $|\trace{O}|\leq \trace{|O|}$, as
\begin{equation}
\label{eq11}
\dot{\ell}_1(\rho_t,\rho_0)\leq \left|\dot{\ell}(\rho_t,\rho_0)\right|\leq \trace{\left|\dot{\rho}_t\right|}=\|\,\dot{\rho}_t\,\|_1\,.
\end{equation}
We immediately conclude that whether we choose the Bures angle \eqref{eq02} or the trace distance \eqref{eq08} only determines the functional dependence of the geometric speed on the choice of the metric \eqref{eq04}. The dynamics and, hence, the actual quantum speed limit, however, is fully characterized by the trace norm of the rate with which the quantum state changes.

We can go even one step further, and consider any Schatten-$p$-norm as a starting point of the derivation,
\begin{equation}
\label{eq12}
\ell_p(\rho_t,\rho_0)=\|\,\rho_t-\rho_0\,\|_p\equiv\left(\trace{\left|\rho_t-\rho_0\right|^p}\right)^{\frac{1}{p}}\,,
\end{equation}
for which the geometric speed is bounded from above by
\begin{equation}
\label{eq13}
\dot{\ell}_p(\rho_t,\rho_0)\leq \|\,\dot{\rho}_t\,\|_p\,.
\end{equation}
A proof of the latter result can be found in \ref{app1}. In conclusion we have that the actual quantum speed limit is given by
\begin{equation}
\label{eq14}
v_\mrm{QSL}\equiv\mi{\|\dot{\rho}_t\|_p \,\,\mrm{for}\,\, p\in[0,\infty)}\,.
\end{equation}
Equation~\eqref{eq14} constitutes our first main result. Derivations of geometric quantum speed limits depend on a rather arbitrary, although well-motivated choice of a measure of distinguishability of quantum states. The final expression will be functionally depended on this choice, see for instance Ref.~\cite{delcampo13} for the relative purity, Ref.~\cite{Deffner2013PRL} for the Bures angle, and Ref.~\cite{Pires2016} for the Wigner-Yanase information. However, since all of these measures fulfill continuity inequalities \cite{Bengtsson2007} (see also Eq.~\eqref{eq08}), the actual quantum speed limit $v_\mrm{QSL}$ is given by the smallest Schatten-$p$-norm of the generator of the dynamics.

\section{Quantum speed in Wigner phase space}

The universal expression of the quantum speed limit, $v_\mrm{QSL}$ in Eq.~\eqref{eq14}, is a powerful expression that can be used to obtain physical insight into the dynamical properties of quantum systems \cite{Cimmarusti2015}. However, computing the tightest bound, i.e., the operator norm \cite{Deffner2013PRL,Simon1979} is far from being a trivial task. Imagine, for instance, we want to study a driven, open quantum system such as in quantum Brownian motion \cite{Hu1992}. In this case, the dynamics is typically solved in a computationally convenient and continuous basis \cite{Ford2001}. Extracting the singular values from such a representation of the time-dependent density operator is computationally expensive, if it is at all feasible. Thus, it would be desirable to find an alternative expression for  $v_\mrm{QSL}$ which gives the same qualitative information, but which is also much easier to compute.

Especially in the treatment of open quantum systems  \cite{Ford2001} as well as to study the semi-classical limit \cite{Dittrich2006} it has proven useful to express quantum states in their Wigner representation
\begin{equation}
\label{eq16}
W(x,p)=\frac{1}{\pi\hbar}\,\int dy\,\bra{x+y}\rho\ket{x-y}\,\ex{-\frac{2 i p \,y}{\hbar}}\,.
\end{equation}
If we want to derive a quantum speed limit in Wigner phase space, we now need to \emph{choose} a measure of distinguishability. To this end, consider the \emph{total variation distance}, which is given by the Wasserstein-1-distance \cite{Wasserstein1969},
\begin{equation}
\label{eq17}
\mc{D}(W_t,W_0)=\|\,W_t-W_0\,\|_1\equiv\int d\Gamma\,\left|W(\Gamma,t)-W_0(\Gamma)\right|\,,
\end{equation}
with $\Gamma=(x,p)$. The Wasserstein-1-distance can be regarded as a generalization of the trace distance to (semi-)probability distributions.

In complete analogy to above, we now consider the geometric speed
\begin{equation}
\label{eq18}
\dot{\mc{D}}(W_t,W_0)=\int d\Gamma\,\frac{W(\Gamma,t)-W_0(\Gamma)}{\left|W(\Gamma,t)-W_0(\Gamma)\right|}\,\dot{W}(\Gamma,t)
\end{equation}
which can again be bounded with the help of the triangle inequality
\begin{equation}
\label{eq19}
\dot{\mc{D}}(W_t,W_0)\leq\left|\dot{\mc{D}}(W_t,W_0)\right|\leq \int d\Gamma\,\left|\dot{W}(\Gamma,t)\right|\,.
\end{equation}
Comparing Eqs.~\eqref{eq11} and \eqref{eq19} we immediately see that we have obtained an analogous expression for the quantum speed limit, $v_\mrm{QSL}$, in Wigner space. In \ref{app2} we show that if we consider the more general case of any Wasserstein-$p$-distance \cite{Wasserstein1969},
\begin{equation}
\label{eq20}
\mc{D}_p(W_t,W_0)=\|\,W_t-W_0\,\|_p\equiv\left(\int d\Gamma\,\left|W(\Gamma,t)-W_0(\Gamma)\right|^p\right)^{1/p}\,,
\end{equation}
we find
\begin{equation}
\label{eq21}
\dot{\mc{D}}_p(W_t,W_0)\leq \|\,\dot{W}_t\,\|_p\,.
\end{equation}
Hence we conclude for the quantum speed limit in phase space, $v^W_\mrm{QSL}$, that we have
\begin{equation}
\label{eq22}
v^W_\mrm{QSL}\equiv\mi{\|\dot{W}_t\|_p \,\,\mrm{for}\,\, p\in[0,\infty)}\,.
\end{equation}
Equation~\eqref{eq22} has the same functional form as the quantum speed limit derived in density operator space \eqref{eq14}. However, Eq.~\eqref{eq22} is significantly easier to compute, since it only involves the absolute value of a real valued function, instead of the singular values of  a high-dimensional operator.

What remains to verify is that $v_\mrm{QSL}$ \eqref{eq14} and $v^W_\mrm{QSL}$ \eqref{eq22} contain the same physical information and that $v_\mrm{QSL}$  and $v^W_\mrm{QSL}$ behave qualitatively similarly.

\section{Qualitative comparison of the two approaches}

In a mathematical sense, the Weyl-Wigner transform \eqref{eq16} is a well-defined, invertible integral transform between the phase-space and operator representations  of quantum states \cite{Schleich2011}. Therefore, one would expect $v_\mrm{QSL}$ \eqref{eq14} and $v^W_\mrm{QSL}$ \eqref{eq22} to be fully equivalent. 

That this is, indeed, the case we will now illustrate by computing $v_\mrm{QSL}$ \eqref{eq14} and $v^W_\mrm{QSL}$ \eqref{eq22} for a solvable example. For the sake of simplicity we restrict ourselves to unitary dynamics induced by the parametric harmonic oscillator with Hamiltonian,
\begin{equation}
\label{eq23}
H=\frac{P^2}{2 M}+\frac{1}{2}M \omega_t^2\,x^2\,.
\end{equation}
It can be shown that the dynamics is fully analytically solvable \cite{Husimi1953}. For systems initially starting in the ground state, $\rho_0=\ket{\psi_0}\bra{\psi_0}$, with
\begin{equation}
\label{eq24}
\psi_0(x)=\left(\frac{M \omega_0}{\pi\hbar}\right)^{1/4}\,\ex{-\frac{m \omega_0 \,x^2}{2\hbar}}
\end{equation}
the time-dependent density operator can be written as \cite{Deffner2013PRE}
\begin{equation}
\label{eq25}
\begin{split}
\rho(x,y,t)&=\sqrt{\frac{M \omega_0}{\pi\hbar}\frac{1}{Y_t^2+\omega_0^2\,X_t^2}}\\
&\quad\times\ex{-\frac{M \omega_0}{2\hbar}\frac{1}{Y_t^2+\omega_0^2\,X_t^2}\left[x^2+y^2+i \left(x^2-y^2\right)\left(\omega_0^2\dot{X}_tX_t+\dot{Y}_tY_t\right)\right]}\,.
\end{split}
\end{equation}
Here, $X_t$ and $Y_t$ are the solutions of the force free harmonic oscillator, $\ddot{X}_t+\omega_t^2 X_t=0$, with the boundary conditions $X_0=0$, $\dot{X}_0=1$, $Y_0=1$, and $\dot{Y}_0=0$.

It is then a simple exercise to numerically obtain the quantum speed limits, $v_\mrm{QSL}$ \eqref{eq14} and $v^W_\mrm{QSL}$ \eqref{eq22}. Without loss of generality, we computed $v_\mrm{QSL}$ \eqref{eq14} from the trace norm \eqref{eq11}, and $v^W_\mrm{QSL}$ \eqref{eq22} from the Wasserstein-1-norm \eqref{eq19}. This is sufficient, since for pure states both the Schatten-$p$-norms as well as the Wasserstein-$p$-norms are monotonic in $p$ \cite{Simon1979}.

\begin{figure}
\includegraphics[width=.49\textwidth]{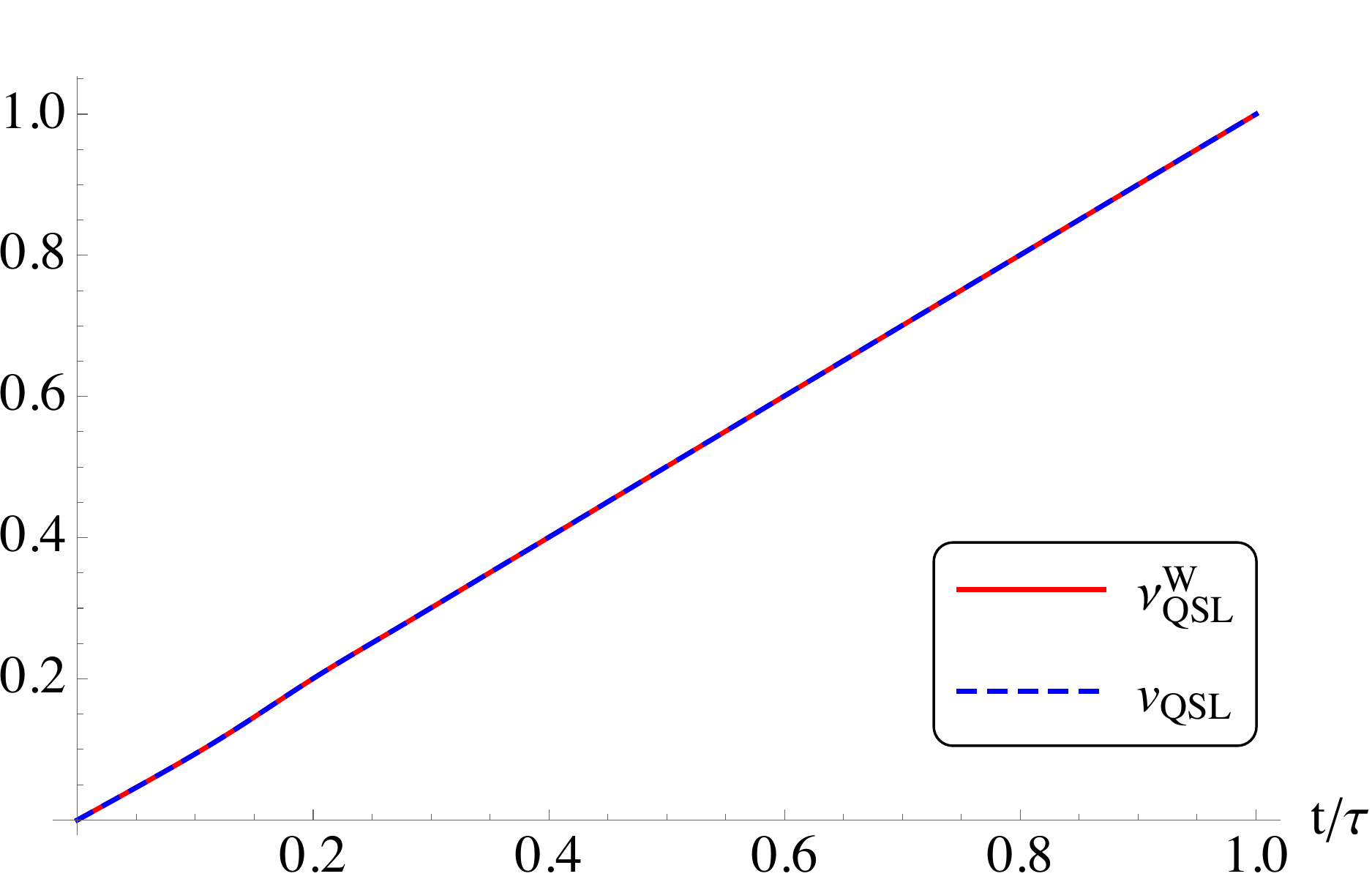}\hfill \includegraphics[width=.49\textwidth]{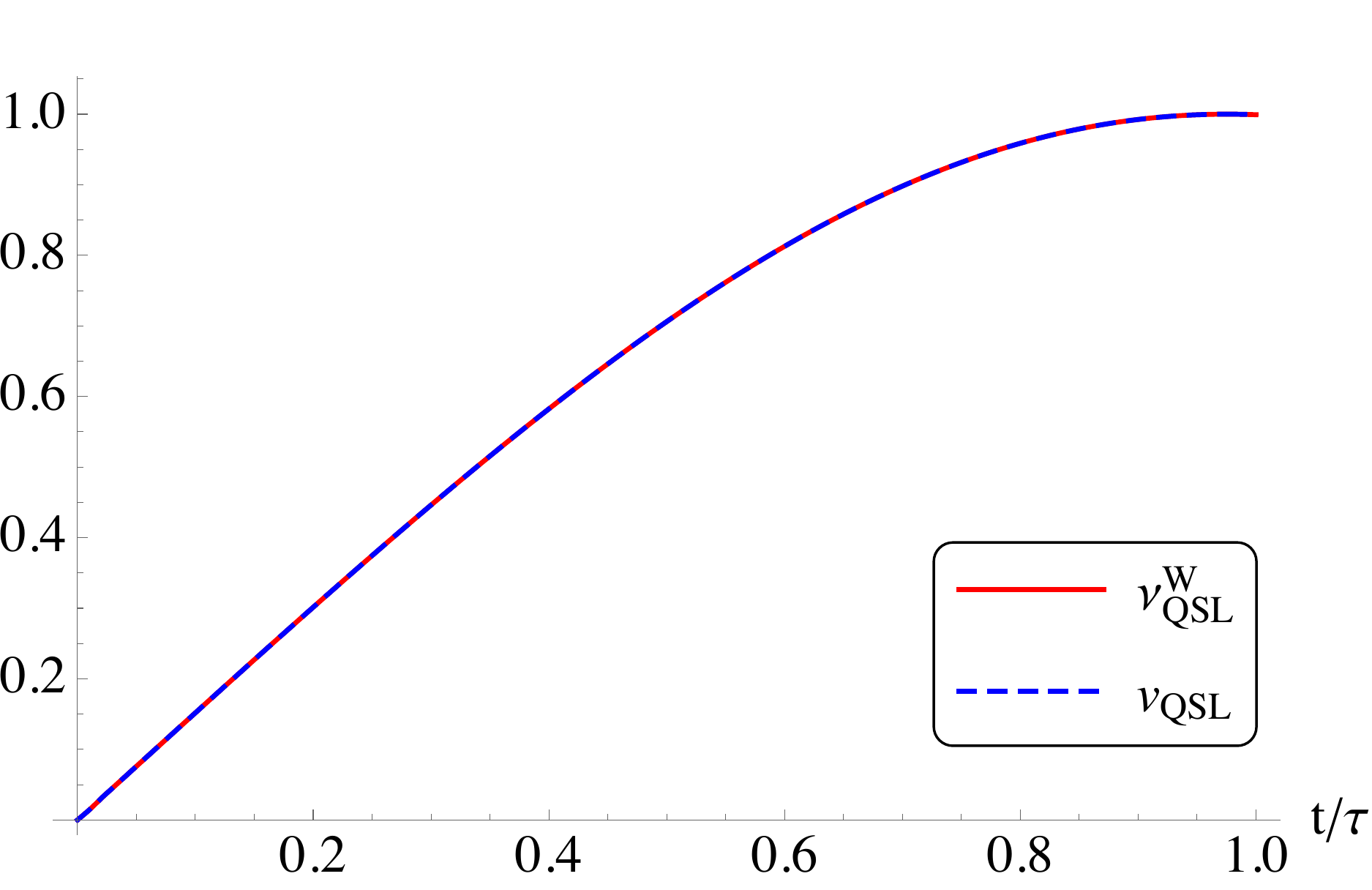}\\
\includegraphics[width=.49\textwidth]{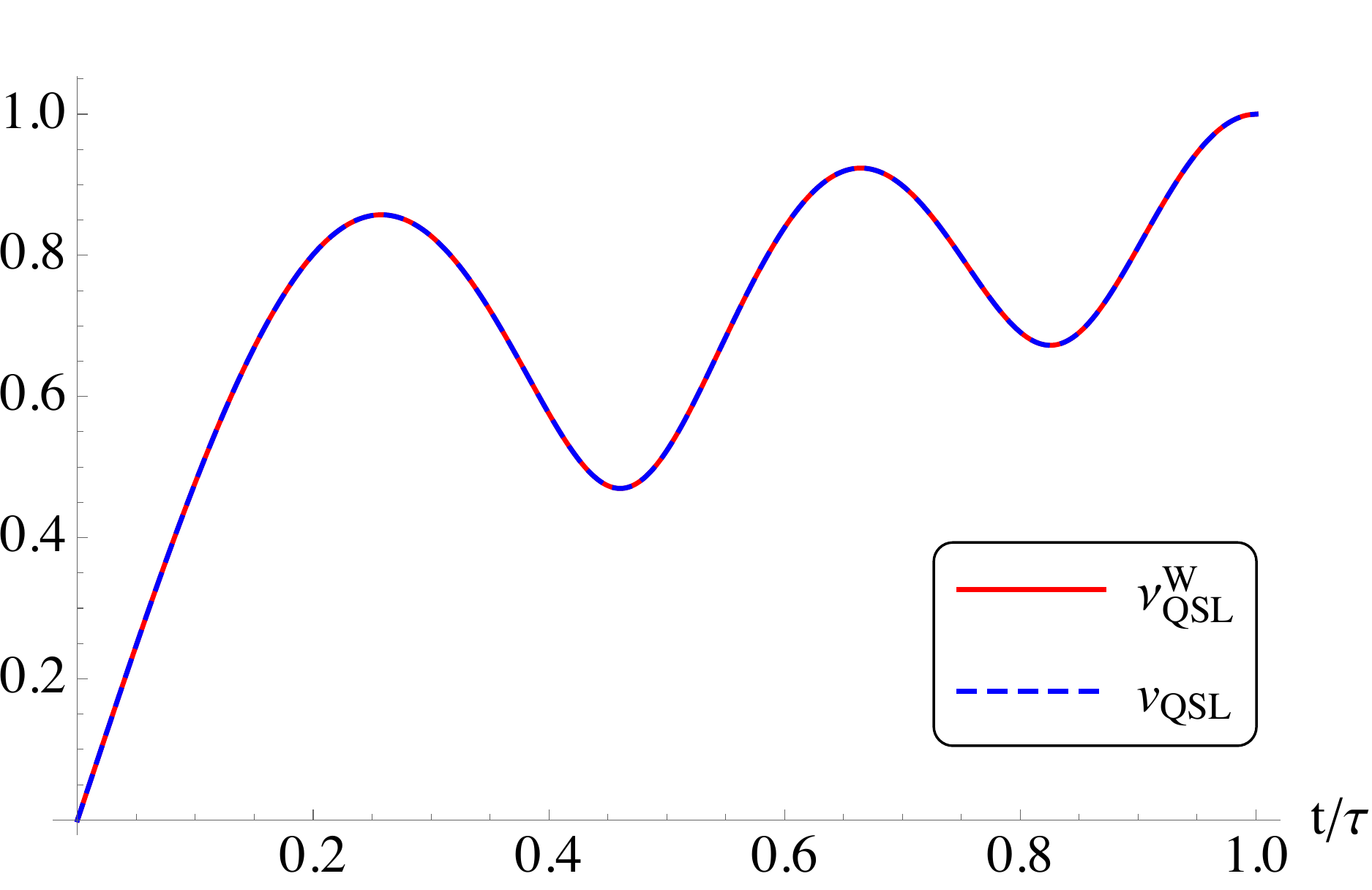}\hfill \includegraphics[width=.49\textwidth]{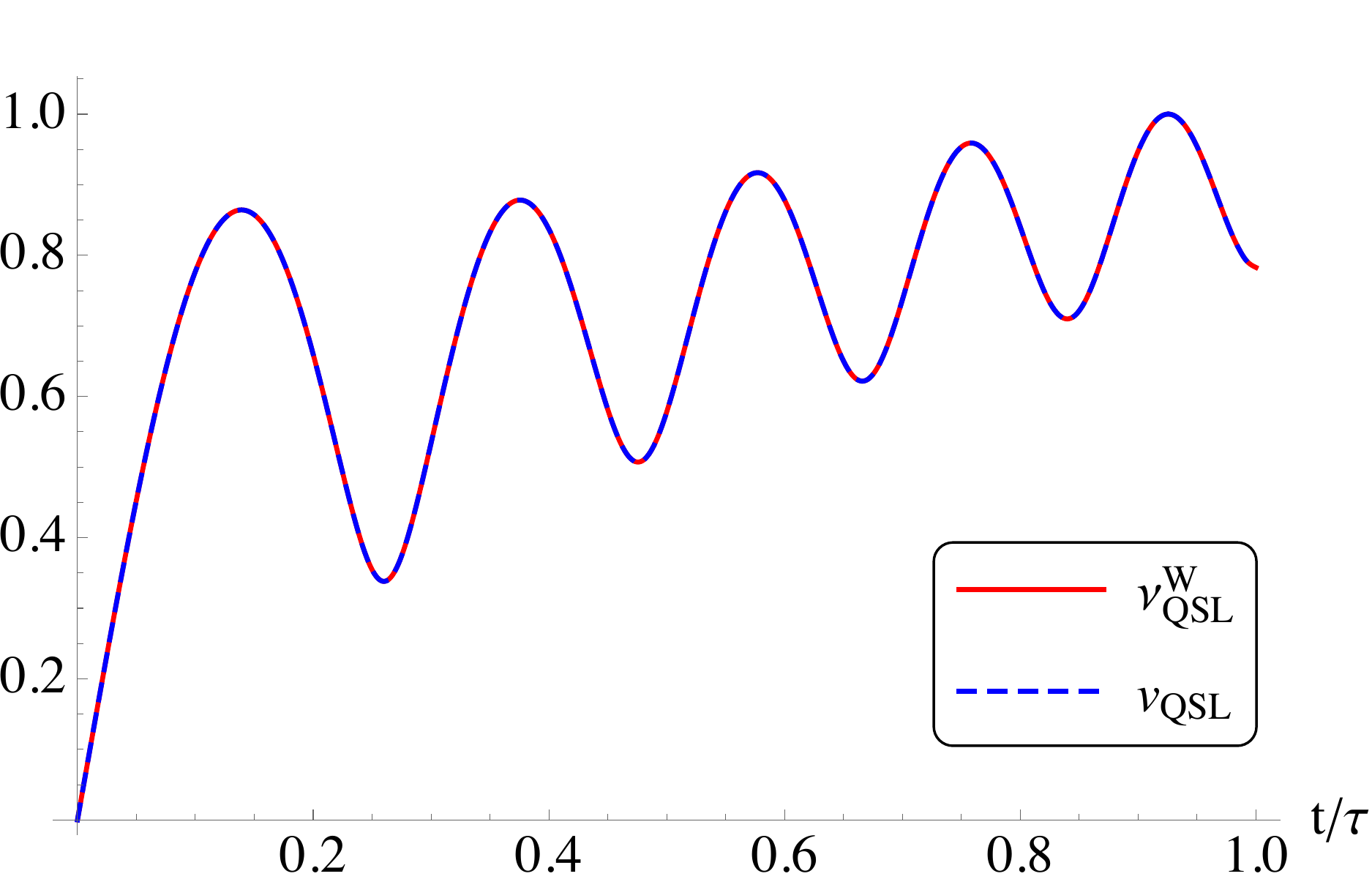}
\caption{\label{fig1} Quantum speed limits in density operator space $v_\mrm{QSL}$ \eqref{eq14} (blue, dashed line) and Wigner phase space $v^W_\mrm{QSL}$ \eqref{eq22} (red, solid line) for $p=1$, the harmonic oscillator \eqref{eq23}, and normalized by their maximal value during the time interval $t\in[0,\tau]$. Parameters are $\omega_0=1$, $\omega_1=2$, $M=1$, $\hbar=1$ and: upper left panel: $\tau=0.1$; upper right panel: $\tau=1$; lower left panel: $\tau=5$; lower right panel: $\tau=10$.}
\end{figure}

Specifically, $v_\mrm{QSL}$ can be obtained from a numerical singular value decomposition of $\rho(x,y,t)$ \eqref{eq25}, which is a computationally rather expensive task. On the other hand, $v^W_\mrm{QSL}$ is obtained directly from the corresponding Wigner function \eqref{eq16}. In Fig.~\ref{fig1} we plot the numerical results for a linear quench, $\omega_t^2=\omega_0^2-(\omega_0^2-\omega_1^2)t/\tau$, for several values of the quench time $\tau$. For the ease of comparison, we further normalized the quantum speed limits $v_\mrm{QSL}$ \eqref{eq14} and $v^W_\mrm{QSL}$ \eqref{eq22} by their maximal value during the time interval $t\in[0,\tau]$. 
We observe perfect agreement of $v_\mrm{QSL}$ \eqref{eq14} and $v^W_\mrm{QSL}$ \eqref{eq22}. Hence, we conclude that our initial expectation is indeed verified, namely we find that $v_\mrm{QSL}$  and $v^W_\mrm{QSL}$ are fully equivalent. In particular for the present case  $v_\mrm{QSL}$  and $v^W_\mrm{QSL}$ only differ by a factors that is determined by their maximal value during the quench time $\tau$.  It is worth emphasizing again that the quantum speed limit is significantly easier to compute in Winger phase space, since the computationally expensive task of having to determine the singular values can be fully avoided.

\section{Quantum speed in the semiclassical limit}

We conclude the analysis by briefly studying the high-temperature, semi-classical limit. Within the geometric approach to quantum speed it often proves useful to define the quantum speed limit time, $\tau_\mrm{QSL}$, which is given by the inverse of the time-averaged quantum speed limit, $v_\mrm{QSL}$, \cite{Deffner2013PRL}. Note that $\tau_\mrm{QSL}$ is \emph{not} a physical time, but rather a charaterisitic of the internal dynamics \cite{mandelstam45,margolus98}.

We can thus define the quantum speed limit time in Wigner space as
\begin{equation}
\label{eq27}
\tau^W_\mrm{QSL}\equiv\frac{\mc{D}(W_t,W_0)}{1/\tau\,\int_0^\tau dt\, v^W_\mrm{QSL}}\,.
\end{equation}
For unitary dynamics it is easy to see that $\tau^W_\mrm{QSL}$ is proportional to $\hbar$, and hence, $\tau^W_\mrm{QSL}$, can be understood as an expression of the Heisenberg indeterminacy principle for energy and time. For open systems, however, the interpretation is less obvious \cite{Deffner2013PRL}. Since the quantum speed limit $v^W_\mrm{QSL}$ is fully determined by the metric properties of the generator of the dynamics \eqref{eq22}, it is not \emph{ad hoc} obvious how $\tau^W_\mrm{QSL}$ behaves in the semiclassical, high-temperature limit $k_B T\gg \hbar\gamma$, where $T$ is the temperature and $\gamma$ the damping coefficient.

To address this question we turn again to the harmonic oscillator. Since we are now interested merely  the behavior of  $\tau^W_\mrm{QSL}$ in the limit high-temperatures, we now consider an un-driven system with potential, $V(x)=1/2\,\,M \omega_0^2\,x^2$. In this case the exact master equation in Wigner space can be written as \cite{Dillenschneider2009,Deffner2013EPL}
\begin{equation}
\label{eq28}
\pd_t\,W(x,P,t)=\left[-\frac{P}{M}\,\pd_x+V'(x)\,\pd_P+\pd_P\left(\gamma P+D_{PP}\,\pd_P\right)+D_{xP}\,\pd^2_{xP}\right]W(x,P,t)
\end{equation}
where $D_{PP}=M\gamma/\beta+M\beta\gamma\hbar^2(\omega_0^2-\gamma^2)/12 $ and $D_{xP}=\beta\gamma\hbar^2/12$. 
\begin{figure}
\includegraphics[width=.49\textwidth]{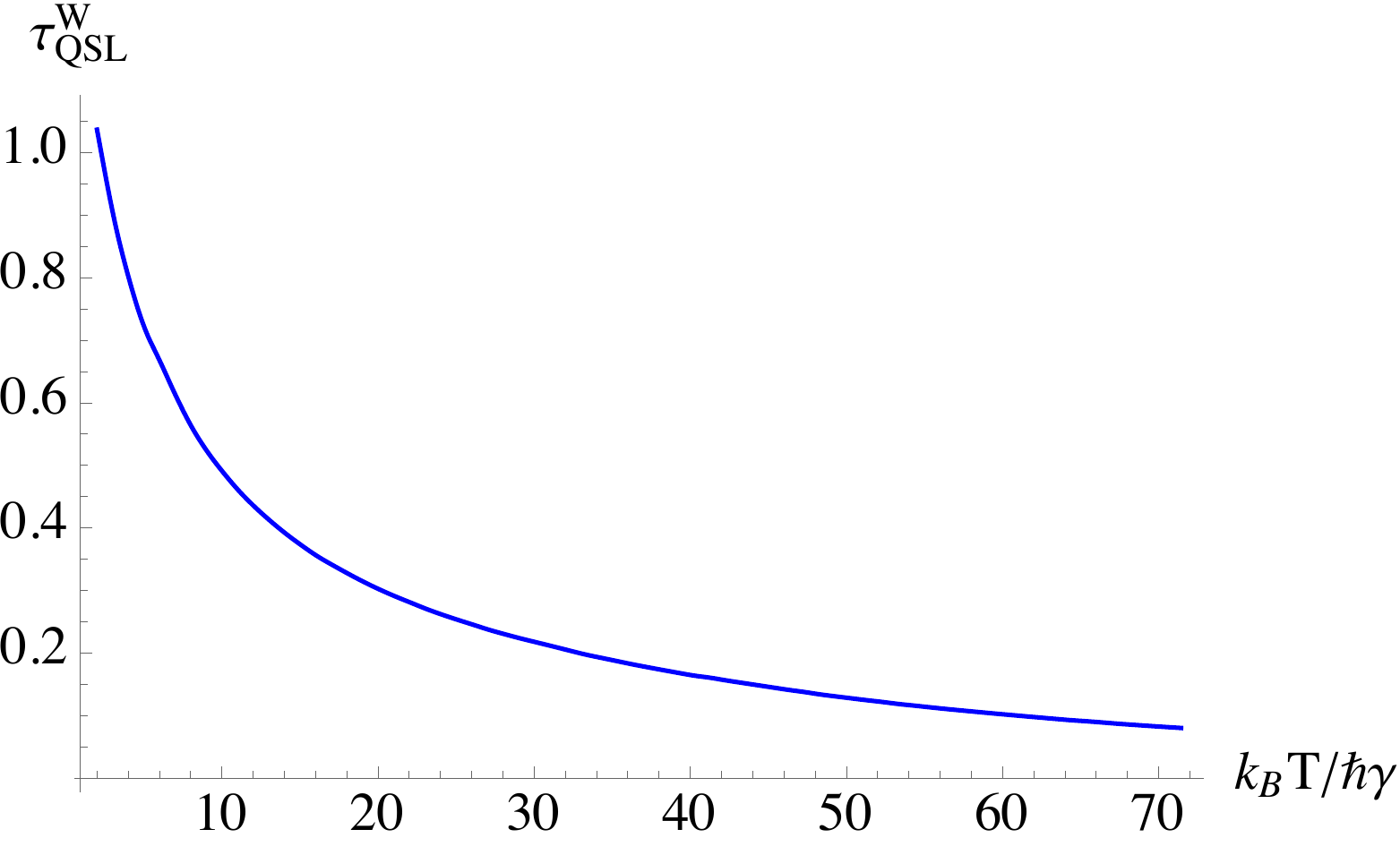}\hfill \includegraphics[width=.49\textwidth]{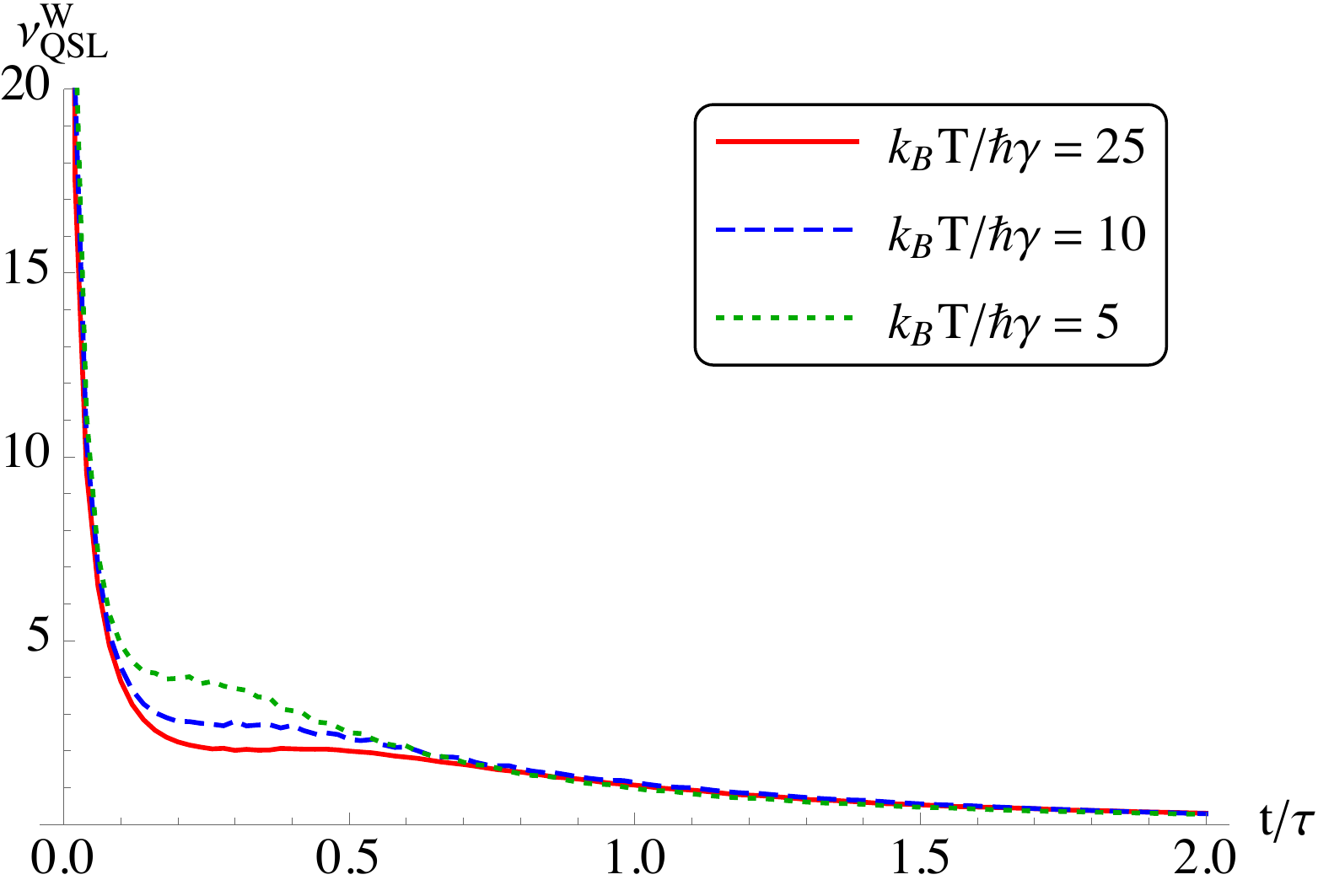}
\caption{\label{fig2} Quantum speed limit time $\tau^W_\mrm{QSL}$ \eqref{eq27} (left panel) and quantum speed limit $v^W_\mrm{QSL}$ \eqref{eq22} for an open harmonic oscillator \eqref{eq28} with the initial state of Eq.~\eqref{eq29} for $p=1$. Parameters are $\gamma=2$, $\hbar=1$, $M=1$, $\omega_0=1$, $\tau=2$, $\mu_x=2$, $\sigma_x=0.5$, $\mu_P=0$, and $\sigma_P=0.5$}
\end{figure}
In Fig.~\ref{fig2} we plot the resulting quantum speed limit, $v^W_\mrm{QSL}$ \eqref{eq22}, together quantum speed limit time, $\tau^W_\mrm{QSL}$ \eqref{eq27}, again for $p=1$. As initial state we chose a narrow Gaussian 
\begin{equation}
\label{eq29}
W_0(x,P)=\frac{1}{2 \pi\, \sigma_x\,\sigma_P}\,\ex{-\frac{(x-\mu_x)^2}{2\sigma_x^2}}\,\ex{-\frac{(P-\mu_P)^2}{2\sigma_P^2}}
\end{equation} 
As expected, the quantum speed limit time $\tau^W_\mrm{QSL}$ vanishes in the classical, high-temperature limit. This observation confirms that the quantum speed limit is a purely quantum phenomenon also in open systems. 

\section{Concluding remarks}

What is the ultimate limit on how fast a quantum system can evolve? At the verge of the age of quantum computing this century old question is more topical than ever. In the present work we highlighted that the maximal quantum speed  can be fully characterized by the Schatten-$p$-norms of the generator of quantum dynamics. We further showed that equivalent expressions can be found in Wigner phase space, where the computationally expensive operator norm is replaced by the absolute value of a real valued function. The utility of the novel approach to quantum speed was illustrated by comparing the quantum speed limits in density operator space and Wigner phase space for the parametric harmonic oscillator. As a consistency check we finally verified that the bound is a pure quantum phenomenon, and that the dynamics of open classical systems is not restricted by the quantum bound. Therefore, we imagine that our results could prove useful for practical consequences and applications of the quantum speed limit, since for any situation the new bound is significantly easier to compute than the operator norm of high-dimensional density matrices.

\ack{The author would like to thank Steve Campbell for insightful discussions. This work was supported by the U.S. National Science Foundation under Grant No. CHE-1648973.}

\appendix

\section{\label{app1}Quantum speed from Schatten distance}

This appendix is dedicated to a proof of Eq.~\eqref{eq13}. Consider the Schatten-$p$-distance
\begin{equation}
\label{a01}
\ell_p(\rho_t,\rho_0)=\|\,\rho_t-\rho_0\,\|_p\equiv\left(\trace{\left|\rho_t-\rho_0\right|^p}\right)^{1/p}\,,
\end{equation}
where $p$ is an arbitrary, positive, real number, $p\in[0,\infty)$. Then the geometric speed can be written as
\begin{equation}
\label{a02}
\dot{\ell}_p(\rho_t,\rho_0)=\left(\trace{\left|\rho_t-\rho_0\right|^p}\right)^{\frac{1}{p}-1}\,\trace{\left[\left(\rho_t-\rho_0\right)^2\right]^{\frac{p}{2}-1}\,\left(\rho_t-\rho_0\right)\,\dot{\rho}_t}\,.
\end{equation}
The latter expression looks rather involved, but it can be simplified by using, $\dot{\ell}_p(\rho_t,\rho_0)\leq |\dot{\ell}_p(\rho_t,\rho_0)| $ and employing the triangle inequality for operators, $|\trace{O}|\leq \trace{|O|}$, to read
\begin{equation}
\label{a03}
\dot{\ell}_p(\rho_t,\rho_0)\leq \left(\trace{\left|\rho_t-\rho_0\right|^p}\right)^{\frac{1}{p}-1} \trace{\left|\rho_t-\rho_0\right|^{p-1} \left|\dot{\rho}_t\right|}\,.
\end{equation}
Equation~\eqref{a03} can be further simplified with the help of H\"older's inequality \cite{Holder1889}
\begin{equation}
\label{a04}
\trace{\left|O B\right|}\leq \left(\trace{\left|O\right|^{q_1}}\right)^{1/q_1}\,\left(\trace{\left|B\right|^{q_2}}\right)^{1/q_2}
\end{equation}
which is true for all $1/q_1+1/q_2=1$. Now choosing $B=\dot{\rho}_t$ and $q_1=p/(p-1)$, for which $q_2=p$, we finally obtain the desired result \eqref{eq13}
\begin{equation}
\label{a05}
\dot{\ell}_p(\rho_t,\rho_0)\leq \|\,\dot{\rho}_t\,\|_p\,.
\end{equation}

\section{\label{app2}Quantum speed from Wasserstein distance}

Finally, we proof the expression for the quantum speed limit in Wigner phase space \eqref{eq22}. To this end, we start with the Wasserstein-$p$-distance \cite{Wasserstein1969}, which is given by
\begin{equation}
\label{b1}
\mc{D}_p(W_t,W_0)=\|\,W_t-W_0\,\|_p\equiv\left(\int d\Gamma\,\left|W(\Gamma,t)-W_0(\Gamma)\right|^p\right)^{1/p}\,.
\end{equation}
Accordingly we have
\begin{equation}
\label{b2}
\begin{split}
\dot{\mc{D}}_p(W_t,W_0)&=\left(\int d\Gamma\,\left|W(\Gamma,t)-W_0(\Gamma)\right|^p\right)^{\frac{1}{p}-1}\\
&\quad\times\int d\Gamma \left|W(\Gamma,t)-W_0(\Gamma)\right|^{p-1}\,\frac{W(\Gamma,t)-W_0(\Gamma)}{\left|W(\Gamma,t)-W_0(\Gamma)\right|}\,\dot{W}(\Gamma,t)\,,
\end{split}
\end{equation}
which can be simplified again with the help of the triangle inequality to read
\begin{equation}
\label{b3}
\begin{split}
&\dot{\mc{D}}_p(W_t,W_0)\leq\left|\dot{\mc{D}}_p(W_t,W_0)\right|\\
&\qquad\leq \left(\int d\Gamma\,\left|W(\Gamma,t)-W_0(\Gamma)\right|^p\right)^{\frac{1}{p}-1}\int d\Gamma \left|W(\Gamma,t)-W_0(\Gamma)\right|^{p-1}\,\left|\dot{W}(\Gamma,t)\right|.
\end{split}
\end{equation}
In complete analogy to the derivation in density operator space, we now consider again H\"older's inequality \cite{Holder1889}
\begin{equation}
\label{b4}
\int dx\left|f(x)g(x)\right|\leq \left(\int dx \left|f(x)\right|^{q_1}\right)^{1/q_1}\,\left(\int dx\left|g(x)\right|^{q_2}\right)^{1/q_2}\,,
\end{equation}
which holds for all $1/q_1+1/q_2=1$. Once again choosing $q_1=p/(p-1)$  and $g=\dot{W}_t $ we finally obtain
\begin{equation}
\label{b5}
\dot{\mc{D}}_p(W_t,W_0)\leq \|\,\dot{W}_t\,\|_p\,.
\end{equation}

\section*{References}

\bibliographystyle{unsrt}
\bibliography{qsl_ps}

\end{document}